\documentclass[useAMS,usenatbib]{mn2e}
\usepackage{graphicx}

\def\gta{\mathbin{\lower 3pt\hbox   
    {$\,\rlap{\raise 5pt\hbox{$\char'076$}}\mathchar"7218\,$}}}

\title[Constraints on Alternatives to Supermassive Black Holes]
{Constraints on Alternatives to Supermassive Black Holes}
\author[M. Coleman Miller]{M. Coleman Miller$^{1}$\\
{E-mail: miller@astro.umd.edu} \\
$^{1}$University of Maryland, Department of Astronomy, College
Park, MD  20742-2421, USA;\\
and Goddard Space Flight Center, Greenbelt,
MD, USA\\}
\begin{document}

\date{Submitted 2005 October}


\maketitle

\label{firstpage}

\begin{abstract}
Observations of the centers of galaxies continue to evolve, and it is
useful to take a fresh look at the constraints that exist on alternatives
to supermassive black holes at their centers.  We discuss constraints
complementary to those of Maoz (1998) and demonstrate that an extremely
wide range of other possibilities can be excluded.  In particular, we
present the new argument that for the
velocity dispersions inferred for many galactic nuclei, even binaries made
of point masses cannot stave off core collapse because hard binaries are
so tight that they merge via emission of gravitational radiation before
they can engage in three-body or four-body interactions. We also show that
under these conditions core collapse  leads inevitably to runaway growth
of a central black hole with a significant fraction of the initial mass,
regardless of the masses of the individual stars.  For clusters of
noninteracting low-mass objects (from low-mass stars to elementary
particles), relaxation of stars and compact objects that pass inside the dark
region will be accelerated by interactions with the dark mass.  If
the dark region is instead a self-supported object such as a fermion ball,
then if stellar-mass black holes exist they will collide with the object,
settle, and consume it. The net result is that the keyhole through which
alternatives to supermassive black holes must pass is substantially
smaller and more contrived than it was even a few years ago.
\end{abstract}


\section{Introduction}

High-resolution observations of the nuclei of many galaxies have
revealed large dark masses in small regions.  These are most
naturally interpreted as supermassive black holes, but as emphasized
by Maoz (1998) it is important to take stock of how rigorously we
can rule out other possibilities.

Here we present arguments
showing that under extremely general conditions almost all
other options are ruled out, further emphasizing that supermassive
black holes are by far the least exotic and most reasonable explanations
for the data in many specific sources.  In \S~2 we lay out our
assumptions, making them as conservative as possible so that
our conclusions are robust.  In \S~3 we show that 
for many observed galactic nuclei, binaries are unable to heat the
stellar distribution effectively because if they are hard then they
merge quickly via gravitational radiation.  This important constraint,
which depends only on dynamics and not the detailed properties of
the specific objects, was not presented by Maoz (1998) or elsewhere
as far as we are aware.  In \S~4 we explore the
consequences of core collapse and demonstrate that a very significant
mass will inevitably coalesce even for point masses.  In \S~5 we
investigate for the first time the consequences if 
stellar-mass black holes exist outside the nucleus.  We show that enough
of them will find their way to the center that they will have serious
effects on the nuclear region, likely consuming a significant amount
of mass and leading to a supermassive black hole.  We discuss the
consequences of this analysis in \S~6.

\section{Assumptions and Dynamics}

In the spirit of Maoz (1998), we make a series of conservative
assumptions to rule out alternatives to supermassive black holes.  Let
us suppose that observations have revealed that a mass $M$ is confined
within a spherically symmetric region whose radius is at most $R$.
We also assume that this mass is composed of identical point
masses $m$; the point mass assumption minimises the interaction between
the masses, and making them identical increases as much as possible the
relaxation time, on which the masses concentrate in the center of the
distribution and hence increase interaction rates. The local two-body
relaxation time for a mass $m$ in a region of mass density $\rho$ and
velocity dispersion $\sigma$ is \citep{Spitzer87}
\begin{equation}
t_{\rm rlx}\approx {1\over{3\ln\Lambda}}{\sigma^3\over{G^2m\rho}}
\label{relax}
\end{equation}
where $\ln\Lambda\sim 10$ is the Coulomb logarithm.  In general this
time depends on radius, but note that if $\rho\sim r^{-3/2}$ and the
velocity dispersion is dominated by a single large mass, the relaxation
time is constant with radius.  

Any streaming motion (e.g., rotation) reduces the relative speed
$\sigma$ and hence reduces the relaxation time (see, e.g.,
\citealt{KLS04} for a numerical treatment of a rotating stellar
system). Therefore, completely random motion leads to the largest
timescales.

For $N$ identical masses in a region 
whose crossing time is $t_{\rm cross}$, the global relaxation time is
approximately \citep{BT87}
\begin{equation}
\begin{array}{rl}
t_{\rm rlx}&\approx {0.14N\over{\ln (0.4N)}}t_{\rm cross}\\
&\approx 10^9~{\rm yr}M_8^{1/2}(1~M_\odot/m)(R/1~{\rm pc})^{3/2}\\
\end{array}
\end{equation}
where $M=10^8M_8\,M_\odot$.
Both expressions for the relaxation time show that for fixed mass
density, lower-mass objects take longer to alter their distribution,
as is expected because two-body relaxation occurs due to graininess
of the gravitational potential, which is less when there are more
objects.

The more concentrated the initial density distribution is, the shorter
will be the central relaxation time (see the extensive discussion in
\citealt{Quinlan96b}).  To be conservative, we therefore
assume a relatively flat distribution such as a Plummer sphere, in
which $\rho\propto (1+r^2/r_c^2)^{-5/2}$, where $r_c$ is the core
radius.  Even for such a distribution, identical point masses will
undergo core collapse within a time (see discussion in \citealt{BT87})
\begin{equation}
t_{cc}\approx 16t_{\rm rlx,h}
\end{equation}
where $t_{\rm rlx,h}$ is the relaxation time at the half-mass radius.
Note that this is a factor $\sim 20$ times less than the time needed
for the cluster to evaporate (\citealt{BT87}). Core collapse of single
objects will formally lead to infinite density at the center.  In
globular clusters and similar systems, this is avoided by the
intervention of binaries: three-body and four-body scattering can
transfer energy from binaries to the stellar velocity dispersion,
heating the cluster and stabilising the density at the center (see
\citealt{Gao91}; \citealt{Fregeau03}; \citealt{GS00} for cluster
simulations involving primordial binaries).  As we now show, however,
when the velocity dispersion is high enough (as it is in many observed
galactic nuclei), binaries cannot prevent core collapse.

\section{The Insufficiency of Binaries}

As shown first by \citet{Heggie75}, binary-single interactions tend to
harden hard binaries, and soften soft binaries.  Only hardening will
inject energy into the cluster and slow core collapse, hence we only
need to consider hard binaries.  For equal-mass objects the hard/soft
boundary is approximately where the orbital energy per object is equal
to the kinetic energy of field stars \citep{Quinlan96a}.  Suppose that
the stellar velocity dispersion is $v_{\rm res}$ at the resolution
radius $r_{\rm res}$ for a particular galactic nucleus.  Then at the
hard/soft boundary the semimajor axis $a$ is given by
\begin{equation}
2Gm/a\approx v_{\rm res}^2\; .
\end{equation}
Any binary emits gravitational radiation as it orbits.  If the time
for the binary to merge by gravitational wave emission is less than
the time for the binary to interact with field stars, then the binary
does not heat the cluster.  For a fixed semimajor axis, the merger
time is maximised for a circular orbit, so we assume $e=0$ to be
conservative.  For comparison, if $e\approx 0.7$ (the mean for a
thermal distribution), the merger time is decreased by a factor $\sim
10$ for fixed $a$.  The rate of change in the semimajor axis from
gravitational radiation, and corresponding merger time for a circular
orbit, is then \citep{Peters64}
\begin{equation}
\begin{array}{rl}
da/dt&=-{64\over 5}G^3\mu m_{\rm bin}^2/(c^5a^3)\\
\tau_{\rm merge}=a/|da/dt|&={5\over 128}c^5a^4/(G^3m^3)\\
&={5\over 8}(c/v_{\rm res})^5 (Gm/v_{\rm res}^3)\\
\end{array}
\end{equation}
where $m_{\rm bin}=m_1+m_2$ is the total mass of the binary, 
$\mu=m_1m_2/m_{\rm bin}$ is the reduced mass, in the
second line we assume $m_1=m_2=m$, and in the third line we
substitute $a=2Gm/v_{\rm res}^2$.

The timescale for a three-body interaction is $\tau_{\rm 3-bod}=
1/(n\Sigma v)$, where $n$ is the number density,  $v\approx
\sqrt{2}v_{\rm res}$ is the relative speed, and  $\Sigma=\pi
r_p^2\left[1+2G(m_{\rm bin}+m)/(r_pv_{\rm res}^2)\right]$ is the
interaction cross section, where $r_p$ is the distance of closest
approach.  For $r_p\approx a$ and three equal masses, a binary at the
hard/soft boundary has $\Sigma\approx 4\pi a^2\approx 16\pi
G^2m^2/v_{\rm res}^4$.   Substituting $n=\rho/m$ we find
\begin{equation}
\tau_{\rm 3-bod}\approx v_{\rm res}^3/\left(16\sqrt{2}\pi\rho G^2m\right)\; .
\end{equation}
The ratio between the merger and three-body timescales is then
\begin{equation}
\tau_{\rm merge}/\tau_{\rm 3-bod}\approx
44(c/v_{\rm res})^5 G^3\rho m^2/v_{\rm res}^6\; .
\end{equation}
This ratio needs to exceed unity for the typical binary to interact
before it merges.  Using the average density $\rho\approx{\bar\rho}=
M/(4\pi R^3/3)$ and assuming a roughly constant velocity dispersion
$v_{\rm res}^2=GM/R$, we find after some manipulation that
$\tau_{\rm merge}/\tau_{\rm 3-bod}>1$ implies
\begin{equation}
\begin{array}{rl}
m&\gta {1\over 3}(v_{\rm res}/c)^{5/2}M\\
&\approx 20\,M_\odot 
v_{\rm res,3}^{5/2}M_8\\
\end{array}
\end{equation}
where $v_{\rm res}=10^3v_{\rm res,3}~{\rm km~s}^{-1}$.
A cluster made of any point masses lighter than this cannot support
itself by binary heating.  

A loophole might appear to be that when there is bulk rotation
(and hence a reduced velocity dispersion) or a density profile in
which the relative speed at the center is much less than
$(GM/R)^{1/2}$, binaries wide enough not to merge quickly could still
heat the distribution.  However, suppose that a binary has tightened
by interactions to the point that $a=2Gm/v_{\rm res}^2$, as
considered above.  Its specific binding energy is then $G\mu/(2a)=
Gm/(4a)=GM/(8R)$, because $v_{\rm res}^2=GM/R$ and thus
$a=2R(m/M)$.  Even if the cluster is 100\% binaries, the total binding
energy liberated by hardening is therefore $GM^2/(8R)$.  The minimum
binding energy of a cluster with mass $M$ and outer radius $R$ is
obtained when all the mass is in a thin spherical shell at radius
$R$, in which case the binding energy is $GM^2/(4R)$.  Even in this 
case, therefore, the maximum effect of binares (prior to their
reaching the previously considered semimajor axis $a=2Gm/v_{\rm res}^2$)
is to increase the cluster binding energy, and hence the cluster
radius, by 50\%.  A smaller binary fraction, a more concentrated cluster,
or nonzero eccentricities for the binaries will all reduce this number.
Therefore, if binaries that are hard relative to $v_{\rm res}$ merge
quickly by gravitational radiation, no possible configuration of
velocities or densities can allow binaries to stall collapse significantly.

\begin{figure}
\includegraphics[width=8cm]{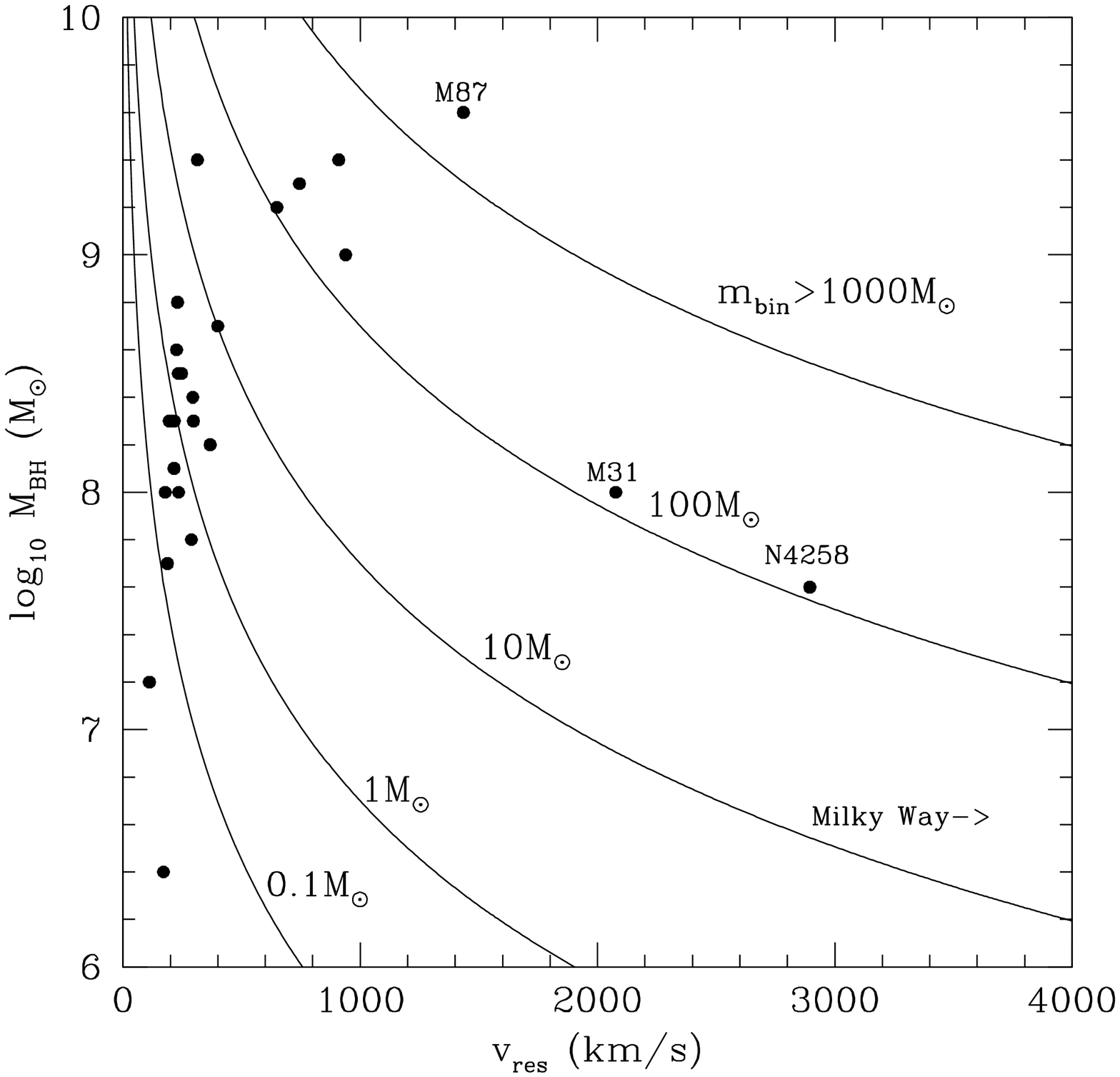}
\caption[fig1]{
Inferred black hole masses and stellar speeds at resolution radius,
derived from  Table~II of \citep{FF05} with updates for M31
\citep{Bender05}, and the Milky Way \citep{Ghez05}, which is far to the right
of the diagram at $v_{\rm res}=1.2\times 10^4$~km~s$^{-1}$.
The curved lines are labeled by the 
minimum mass of identical point masses such that if they make up the dark
mass, binaries can in principle heat the system and delay core
collapse.  Several galaxies have  $M_{\rm
min}>100\,M_\odot$ (the Galaxy has $M_{\rm min}\approx 400\,M_\odot$)
and hence no reasonable stellar component could
heat the system.
}
\label{binary}
\end{figure}

Figure~\ref{binary} plots the black hole
mass versus the stellar velocity at the resolution radius, along 
with the minimum mass of point masses that would allow binary heating.
Several galactic nuclei cannot be heated by masses lower than $100\,M_\odot$,
including the Galaxy, M87, M31, and NGC~4258.  If such masses were
assembled, the number of objects would therefore be small for
a given dark mass, which would reduce the relaxation time dramatically
(see equation~\ref{relax}) and would mean that the evaporation time
$t_{\rm evap}\approx 300~t_{\rm rlx}$ would be much less than a Hubble
time.  Therefore, even with an implausible collection of $>100\,M_\odot$
objects in binaries, the cluster would still disintegrate rapidly.

\section{Core Collapse}

If core collapse happens, what is the result?  \citet{Cohn80} found that the
density profile approaches $n\propto r^{-2.23}$.  For ease of
calculation, and to be conservative, we will assume a shallower
profile $n\propto r^{-2}$, which is appropriate for a singular
isothermal sphere.  In such a profile, the total mass interior
to radius $r$ is proportional to $r$, and the velocity dispersion
is constant with radius.

In the high density central regions, even point masses can merge
because they emit gravitational radiation.  \citet{QS89}
showed that for a relative speed $v$ at infinity between two masses
with reduced mass $\mu$ and total mass $m_{\rm tot}$, there will be
a mutual capture if the pericenter distance of approach $r_p$ satisfies
\begin{equation}
r_p<r_{p,{\rm max}}=\left(85\pi\sqrt{2}\over{12 c^5}\right)^{2/7}
G\mu^{2/7}m_{\rm tot}^{5/7}v^{-4/7}\; .
\end{equation}
For equal masses and $v\approx \sqrt{2}v_{\rm res}$, the cross section for 
merging in the gravitationally focused limit is
\begin{equation}
\Sigma_{\rm merge}\approx 2\pi r_{p,{\rm max}}(Gm_{\rm tot}/v^2)
\approx 19\left(Gm\over c^2\right)^2\left(c\over v_{\rm res}\right)^{18/7}\; .
\end{equation}
Over a time $T$, the probability of merger of an average point mass is then 
$P=Tn\Sigma v_{\rm res}$.  The average number density is 
${\bar n}=(M/m)/(4\pi R^3/3)$.  At this density, we find after some
algebra that the probability is
\begin{equation}
{\bar P}\approx 4T{m\over M}\left(v_{\rm res}\over c\right)^{10/7}
{v_{\rm res}^3\over{GM}}\; .
\end{equation}
With the rough approximation that $n\approx {\bar n}(r/R)^{-2}$ and
$M(<r)\approx (r/R)M$, this implies that the enclosed mass $M_{\rm merge}$
inside of which the masses merge in time $T=10^9T_9$~Gyr is
\begin{equation}
\begin{array}{rl}
M_{\rm merge}&\approx {\bar P}^{1/2}M\approx 3(c^3/G)^{1/2}
T^{1/2}m^{1/2}(v_{\rm res}/c)^{31/14}\\
&\approx 5\times 10^5\,M_\odot T_9^{1/2}
(m/1\,M_\odot)^{1/2}v_{\rm res,3}^{31/14}\; ,\\
\end{array}
\end{equation}
or just $M$ if ${\bar P}>1$.
The net result is that even for low-mass objects, core collapse will
lead to the formation of a large single mass at the center of the
distribution.  However, as is clear from Equation~(\ref{relax}), if the
component masses are small enough then the relaxation time is so large
that core collapse will not occur.  We now address this situation.

\section{Dynamical Friction and Stellar-Mass Black Holes}

Suppose that the particles comprising the matter are very low-mass
indeed, such as elementary particles.  Suppose also that, like
hypothesised dark matter, the particles interact neither with 
themselves nor with ordinary baryonic matter in any way but
gravitationally.  If in some improbable circumstance the particles
have collected in a cluster of total mass $M$ and radius $R$, what
will affect them?

Because the particles have low mass, any more massive
objects that enter their region will sink to the center via dynamical
friction.  The characteristic time for a mass $m$ to sink is 
(see \citealt{BT87} for a discussion)
\begin{equation}
\tau_{DF}\approx v_M^3/\left[4\pi\xi\ln\Lambda G^2\rho m\right]
\end{equation}
where $v_M$ is the speed of the massive object, $\ln\Lambda$ is a
Coulomb logarithm, and
\begin{equation}
\xi={\rm erf}(X)-{2X\over{\sqrt{\pi}}}e^{-X^2}
\end{equation}
with $X\equiv v_M/(\sqrt{2}\sigma)$.  If $v_M\approx v_{\rm res}$, then
$\xi\approx 0.2$.  Adopting as before $v_{\rm res}\approx (GM/R)^{1/2}$
we find
\begin{equation}
\begin{array}{rl}
\tau_{DF}&\sim 0.2(M/m)(GM/v_{\rm res}^3) \\
&\approx 8\times 10^9~{\rm yr}M_8^2(1~M_\odot/m)
v_{\rm res,3}^{-3}\; .\\
\end{array}
\end{equation}
This implies that for systems such as the central region of M31,
where $v_{\rm res}\approx 2000$~km~s$^{-1}$ and $M>10^8\,M_\odot$ 
\citep{Bender05},
even ordinary stars will sink to the center of the mass distribution
within a few Gyr, or much less if the dark matter is more
concentrated.  Therefore, all ordinary stellar processes that would
proceed around a supermassive black hole will also proceed around a
concentrated region of noninteracting particles, except that stars
inside the region will sink to the center rapidly (see
\citealt{Quinlan96b}).  Thus if the dynamical friction time at the
average density ${\bar\rho}=M/(4\pi R^3/3)$ is less than a few Gyr,
stars and compact objects that enter the region will collide, merge,
and have prime conditions for forming a large single mass.

The rate of interactions of stars with the central concentrated
region is less for smaller regions. Suppose that the
non-stellar matter is very concentrated, say with a radius just a
few times the radius of a black hole with the same mass.  Then, the
arguments used to estimate rates of extreme mass ratio inspirals
also apply here. These arguments suggest that stellar-mass black
holes will spiral into supermassive black holes at a rate not less
than
$\sim 10^{-8}$~yr$^{-1}$ 
\citep{HB95,SR97,MEG00,Freitag01,Freitag03,Ivanov02,HA05}.  
Therefore, regardless of
how compactly the dark matter is distributed, if stellar-mass black 
holes exist they will enter the mass distribution in much less
than a Hubble time.

The mass accreted by a black hole during inspiral is comparatively
small.  For example, consider a constant-density region 
$\rho={\bar\rho}=M/\left({4\over 3}\pi R^3\right)$ with nonrelativistic
particles moving at an average speed $v_{\rm res}=(GM/R)^{1/2}$
relative to the black hole.
The cross section for absorption by a black hole of mass $m$ is
$\Sigma=(4Gm/c^2)(2Gm/v^2_{\rm res})$, so during a time $\tau_{DF}$
the black hole will accrete a mass
\begin{equation}
\begin{array}{rl}
\Delta m&=\rho\Sigma v_{\rm res}\tau_{DF}\\
&\approx 0.4\left(v_{\rm res}/c\right)^2m\; .\\
\end{array}
\end{equation}
This is therefore only a small fraction of the original mass.  
Similarly, if after inspiral the black hole is fixed at the center
of the mass distribution, it accretes little mass.

This conclusion changes if the black hole wanders freely around
the dark matter distribution.  This could happen if, for example,
multiple massive objects enter the dark matter region and scatter
each other frequently.  In this case, for the same assumptions as
before, the mass accretion rate ${\dot m}=\rho\Sigma v_{\rm res}$ 
becomes
\begin{equation}
{\dot m}=2{m^2\over M^2}{\sigma^5\over{Gc^2}}\; ,
\end{equation}
implying a growth time
\begin{equation}
\begin{array}{rl}
T_{\rm growth}&={1\over 2}(M/m)\left(GMc^2/\sigma^5\right)\\
&\approx 2\times 10^{14}~{\rm yr}M_8^2
(m/10\,M_\odot)^{-1}v_{\rm res,3}^{-5}\; .\\
\end{array}
\end{equation}
This is not constraining on most supermassive black hole candidates, 
but for the Galaxy ($M\approx 4\times 10^6\,M_\odot$; \citealt{Ghez05}) and
$v_{\rm res}\approx 1.2\times 10^4$~km~s$^{-1}$ within 45~AU 
\citep{Ghez05}, the
growth time is $T_{\rm growth}\sim 4\times 10^5$~yr.  Radio observations
\citep{RB04,Shen05} suggest that at least $4\times 10^5\,M_\odot$ is
contained within $\sim 0.5$~AU of the position of Sgr~A$^*$, which
lowers the accretion time to at most $\sim 100$~yr.  Note that a doubling
of mass decreases the time to the next doubling by a factor of
two, so substantial growth results in a runaway.  Note also that because
by assumption the only matter entering the black hole does so
with prompt infall and without release of radiation, the growth
is not limited by the Eddington rate.  If the particles are baryonic
or otherwise have a reasonable strength of self-interaction, then the
accretion rate is greatly enhanced, up to a possible Eddington-like
maximum.

Finally, suppose that the non-luminous matter is in fact in the
form of a star supported by pressure gradients rather than by
simple motion as we have assumed up to this point.  An example
would be fermion balls (see, e.g., \citealt{TV98}), which are
collections of massive neutrinos supported by degeneracy pressure.
In that case, clearly a stellar-mass black hole captured by the
star will consume matter at the star's center and remove pressure
support, leading to rapid destruction of the star.

\section{Conclusions}

In the past decade, thanks to many observational developments, the case
for supermassive black holes in the centers of many galaxies has gone
from strong to essentially inescapable.   We have shown that for many
specific galactic nuclei, the observational constraints are strong
enough to rule out binary heating, hence the relevant evolution time is
the time to core collapse.  This is a factor of $\sim 20$ less than the
time to evaporation, which has previously been used as the conservative
standard for stellar cluster persistence. For many individual galactic
nuclei, therefore, the combination of time to core collapse
and lack of binary heating rules out dense stellar
clusters as an alternate explanation for the inferred dark mass.  
Specifically, the Galaxy, NGC~4208, and M31 have core collapse
times $<$2~Gyr for $0.5\,M_\odot$ objects and cannot be stabilised
by binaries less than $100\,M_\odot$.  M32 also has a core collapse
time $<$2~Gyr, but could in principle be stabilised by stellar-mass binaries.
All other sources currently have core collapse times $>$200~Gyr
for $0.5\,M_\odot$ objects.

The only remaining possibilities are concentrated regions of noninteracting
low-mass particles or self-supported exotic objects such as a fermion
balls \citep{TV98}.  Even in this case, we have shown that dynamical
evolution of the stars and black holes near the centers of galaxies will
cause multiple stellar-mass black holes to fall to the center of the
potential, if black holes exist at all.  Such black holes would consume
any high-mass exotic pressure-supported objects, and would also accrete
a noninteracting cluster of particles if allowed to move around freely.
Therefore, the existence of stellar-mass black holes would lead to the
production of supermassive black holes in many specific sources even if
the supermassive holes did not form in other ways.  When combined with
the high redshifts inferred from Fe~K$\alpha$ lines in some Seyfert
galaxies \citep{RN03}, dramatic deviations from standard physics are
required to explain observations in ways not involving black holes.

\section*{Acknowledgments}

We thank Laura Ferrarese, Marc Freitag, and Doug Richstone for valuable
discussions.  This work was supported in part by NASA ATP grant NAG
5-13229, and by a senior NRC fellowship during a sabbatical at Goddard
Space Flight Center.

\end{document}